\documentclass[twocolumn,amsmath,amssymb]{revtex4}

\usepackage{graphicx}
\usepackage{dcolumn}

\begin{document}
\title{Experimental quantum key distribution with active phase randomization}
\author{Yi Zhao, Bing Qi, and Hoi-Kwong Lo}
\affiliation{Center for Quantum Information and Quantum Control,
Department of Physics and Department of Electrical \& Computer
Engineering, University of Toronto, Toronto, Ontario, M5S 3G4,
Canada}
\begin{abstract}
Phase randomization is an important assumption made in many security
proofs of practical quantum key distribution (QKD) systems. Here, we
present the first experimental demonstration of QKD with reliable
active phase randomization. One key contribution is a
polarization-insensitive phase modulator, which we added to a
commercial phase-coding QKD system to randomize the global phase of
each bit. We also proposed a simple but useful method to verify
experimentally that the phase is indeed randomized. Our result shows
very low QBER ($<1\%$). We expect this active phase randomization
process to be a standard part in future QKD set-ups due to its
significance and feasibility.
\end{abstract}
\maketitle

Quantum key distribution (QKD) allows two authenticated parties
 to share a secret key
\cite{BB84,Gisin2002}. The security of this secret key has been
rigourously proven to be unconditional \cite{securityproof}.
Experimentally, people have demonstrated QKD over 175km optical
fibre \cite{longdistance}. Moreover, commercial QKD systems are now
available \cite{commercialQKD}.

How secure are these implementations of QKD? The security proofs
\cite{securityproof} rely on several assumptions. A frequently used
one is that Alice (the sender) has a perfect single photon source.
However, most QKD experiments use heavily attenuated coherent laser
sources due to the great challenge to build a perfect single photon
source \cite{ExpQKD,ContQKD,stuck2002}. This substitution causes
some security concern (like the photon-number-splitting attack
\cite{PNS}) though, the security for QKD using weak coherent state
is still provable \cite{ILM,GLLP}. Recent study shows that by
introducing decoy states, the signals can be rather strong without
jeopardizing the security \cite{HwangDecoy,LoDecoy,OtherDecoy}. QKD
with decoy states has been experimentally demonstrated recently
\cite{ExpDecoy}.

The assumption of single photon source was removed at the price of
introducing another assumption: the phase of the quantum signal is
uniformly random \cite{ILM,GLLP,HwangDecoy,LoDecoy,OtherDecoy}, and
thus is inaccessible to the eavesdropper. It has been shown that
phase randomization of the quantum signal is a crucial security
requirement rather than a tricky assumption \cite{LoPhase}.
The existing security proofs of non-randomized phase QKD
\cite{NonRandomizePhase} are all at a price of comprising
performance. Particularly, they cannot be applied to the security
proof of decoy state QKD.

QKD experiments with intentionally randomized phase have never been
reported. Here we remark that the quantum signals sent by Alice are
not ``naturally'' phase-randomized. For example, in uni-directional
QKD system, strong ancillary pulses (sometimes called reference
pulses) are often used for feed-back control to stabilize the
asymmetric Mach-Zehnder interferometer (MZI) \cite{ContQKD}. The
phase of such strong classical pulses could be (in principle)
accurately measured, leaking the phase information to the
eavesdropper. Even if weak signals are used, the phase coherence of
the laser source could maintain for many emissions of weak signals,
which makes it possible to measure the phase accurately. For
bi-directional system (``plug \& play'' system) \cite{stuck2002},
classical pulses sent from Bob cause the same problem as the strong
ancillary pulses in uni-directional system do.

Active phase randomization in real QKD system is challenging: the
phase modulator should be polarization insensitive; this extra phase
modulator has to be carefully synchronized with the original system
to randomize the phase in real-time; since the output from Alice is
very weak ($\sim0.1$ photon per pulse), it is not straightforward
how to verify that the phase is indeed randomized; the phase
randomization must not increase the quantum bit error rate (QBER)
significantly.

In this paper, we will present the first QKD experiment with
reliable active phase randomization. Our implementation is based on
a modified commercial ``plug \& play'' system \cite{stuck2002}. The
global phase of each bit is randomized by an additional phase
modulator, after which the bit is sent to Bob. This phase modulator
is designed to be polarization-insensitive. Therefore neither
polarization-maintaining fiber nor dynamic polarization control is
necessary. Our result shows that the phase difference between
adjacent signals has been confidently randomized by this
phase-randomization phase modulator, while the relative phase
between the two pulses of the same signal is solely determined by
Alice's coding phase modulator. We expect phase randomization to
become a standard part in future QKD systems due to its security
significance \cite{ILM, GLLP, LoPhase} and feasibility shown in this
paper.

\begin{figure}[!t]
  \includegraphics[width=6.5cm]{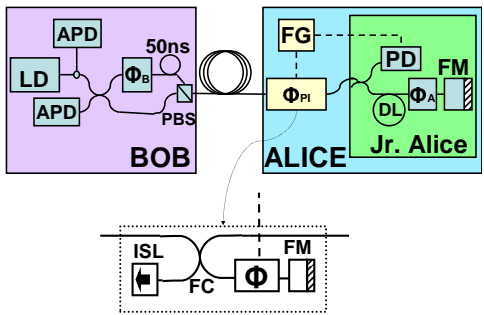}
  \caption{Upper Chart: Schematic of the experimental
set-up in our system. Inside Bob/Jr. Alice: components in
Bob/Alice's package of id Quantique QKD system. Our modifications:
$\Phi_{\mathrm{PI}}$: polarization-insensitive phase modulator
(detailed structure shown in lower chart); FG: functional generator.
Original QKD system: LD: laser diode; APD: avalanche photon diode;
$\Phi_\mathrm{i}$: phase modulator; PBS: polarization beam splitter;
PD: classical photo detector; FM: faraday mirror. Lower chart:
Detailed structure of the polarization-insensitive phase modulator.
ISL: optical isolator; FC: 2$\times$2 Fiber Coupler; $\Phi$:
electro-optical modulator; FM: faraday mirror. Solid line: SMF28
single mode optical fiber; dashed line: electric cable.
}\label{Fig:Schematic}
\end{figure}

The schematic of our set-up is shown in FIG. \ref{Fig:Schematic}.
The original QKD system works as follows: Bob generates a frame of
laser pulses at 5MHz repetition rate; each pulse is splitted into two by the asymmetric MZI;
the one propagates through the shorter arm is called the reference
pulse, and the one propagates through the longer arm is called the
signal pulse; the insertion loss of the phase modulator
($\Phi_\mathrm{B}$ in FIG. \ref{Fig:Schematic}) makes the signal
pulse weaker than the reference pulse. The state of the $i$th bit
emitted from Bob is $|\alpha_i\rangle|\beta_i\rangle$, where
$|\alpha_i\rangle$ denotes the reference pulse and $|\beta_i\rangle$
denotes the single pulse. Pulses are sent to Alice 
and are reflected by the faraday mirror (FM in FIG.
\ref{Fig:Schematic}, upper chart); Alice encodes the quantum
information by modulating the phase of the signal pulse with her
phase modulator ($\Phi_\mathrm{A}$ in FIG. \ref{Fig:Schematic}); the
pulses are then attenuated to single photon level and sent back to
Bob. The state sent from Alice to Bob is $|\alpha_i'\rangle|\beta_i'
e^{i\phi_{Ai}}\rangle$. Bob decodes the quantum information by
modulating the phase of the reference pulse with his phase modulator
($\Phi_\mathrm{B}$ in FIG. \ref{Fig:Schematic}) and letting the two
pulses interfere at the
coupler before sending the next frame of pulses.

\begin{figure}[!t]
  \includegraphics[width=6.5cm]{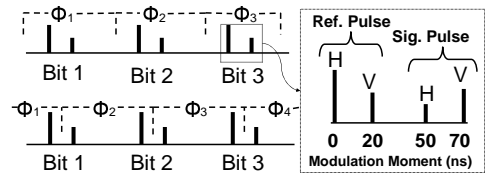}\\
  \caption{Upper chart: correctly implemented phase randomization.
  A different random phase is applied on each bit. Lower chart:
phase modulation for incorrect implementation: signal pulse and
reference pulse are modulated by different phase and thus we could
expect QBER to be around 50\%. Right Chart: differently polarized
components of the same pulse are modulated at different
moments.}\label{Fig:Pulse_and_Phase}
\end{figure}

Alice should modulate the global phase of each signal with an extra
random value to implement phase randomization. i.e., the state
emitted from Alice should be $|\alpha_i 'e^{i\phi_i}\rangle|\beta_i'
e^{i(\phi_i+\phi_{Ai})}\rangle$, where $\phi_i$ should be a random
value for each bit as shown in FIG. \ref{Fig:Pulse_and_Phase} (upper
chart). The birefringence of optical fiber makes the polarization of
laser unpredictable and changing frequently. Therefore the phase
modulator should be polarization insensitive.

There have been several proposals on polarization-insensitive phase
modulators, based either on liquid crystal (LC) \cite{LC_PIPM} or on
acousto-optic modulator (AOM) \cite{AOM_PIPM}. LC-based phase
 modulators require sophisticated fabrication, and AOM-based phase
 modulators cannot meet the repetition rate of the laser source (5MHz).
 Therefore we need to design
 another polarization-insensitive phase modulator, which consists of
 commercial parts and can work at several megahertz.

Our design of phase modulator is shown in FIG. \ref{Fig:Schematic}
(lower chart). It can be easily shown that the phases of both
vertically- and horizontally-polarized components of incoming light
are modulated by propagating through the phase modulator ($\Phi$ in
FIG. \ref{Fig:Schematic}, lower chart) twice and the $\pi/2$
polarization rotation due to the faraday mirror (FM in FIG.
\ref{Fig:Schematic}, lower chart).

The synchronization signal from the photo detector (PD in FIG.
\ref{Fig:Schematic}) will trigger the functional generator (FG in
FIG. \ref{Fig:Schematic}) when the pulse frame enters Alice. The
functional generator will hold for a time period before outputting a
pre-loaded uniformly random voltage pattern (generated by id
Quantique quantum random number generator) to drive the
polarization-insensitive phase modulator to randomize the phase of
each bit. This phase modulation extends to the full range of
$[0,2\pi]$ with amplitude resolution of 12 bits as limited by the
functional generator. The linearity of electro-optical effect
\cite{YarivBook} guarantees that the phase applied on each bit is
also uniformly random. In our set-up, the frame length is 504
pulses.

This phase randomization process does not affect the performance of
QKD system. The phase modulator applies the same phase shift to both pulses of the same bit as shown
in FIG. \ref{Fig:Pulse_and_Phase} (upper chart). Therefore the
QBER should not change. Moreover,
since this extra phase modulator is in Alice's side, we can set the
output intensity from Alice arbitrarily. Thus it does not affect the
gain. This is good news because the QKD system will not pay any
price on performance for randomizing the phase. However, it leaves
us a problem: how can we see that the phase is reliably randomized?

Our answer is to shift the delay time of the functional generator
(FG in FIG. \ref{Fig:Schematic}) so that the two pulses of the same
bit are modulated differently, as shown in FIG.
\ref{Fig:Pulse_and_Phase} (lower chart). The relative phase between
the two pulses is then uniformly random, and we should observe a
sharp increase of QBER to around 50\%.

\begin{figure}
  \includegraphics[width=6cm]{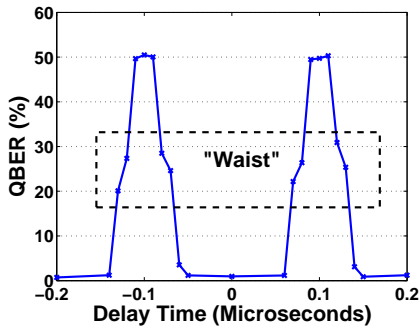}\\
  \caption{QBER versus delay time. Alice sent 843k bits with 0.1 photon
per bit on average for each point in this diagram. QBER at t=0$\mu$s
is surprisingly low ($<$1\%), and is comparable to QBER at
t=$\pm$0.2$\mu$s.}\label{Fig:QBER}
\end{figure}

We shift the delay time for $\pm0.2\mu$s, i.e., a range of two
periods of the QKD system, at a step of 10ns (larger step is used
for the flat area). The result is shown in FIG. \ref{Fig:QBER}. We
can see clearly that when the two pulses of the same bit are
modulated equally as in FIG. \ref{Fig:Pulse_and_Phase} (upper
chart), the QBER is indeed low ($<1\%$), as the central flat part in
FIG. \ref{Fig:QBER}. This confirms our prediction of low-QBER.
However, when the delay time is shifted to the
value that the two pulses are modulated differently (as in FIG.
\ref{Fig:Pulse_and_Phase}, lower chart), the QBER would increase to
around 50\% as the two spikes in FIG. \ref{Fig:QBER} show. The
fluctuation is only $\pm0.5\%$ (which is within one standard
deviation) from the expected value  of $50\%$.

We surprisingly found a ``waist'' on each slope of the spikes,
making the slopes neither sharp nor smooth. The explanation we found
is that the vertically- and the horizontally-polarized components of
a pulse are modulated at different time: one is modulated when the
pulse propagates toward the faraday mirror (FM in FIG.
\ref{Fig:Schematic}, lower chart), and the other one is modulated
when the pulse is reflected back as shown in FIG.
\ref{Fig:Pulse_and_Phase} (right chart). This time difference makes
it possible that the phase modulation applied on the two components
are different when we shift the modulation time gradually. This
modulation difference will increase the QBER to a value between a
few percents to 50\%, depending on the polarization of the pulse.

The fiber connecting the phase modulator ($\Phi$ in FIG.
\ref{Fig:Schematic}, lower chart) and the faraday mirror (FM in FIG.
\ref{Fig:Schematic}, lower chart) is roughly 2m. Therefore the time
difference between the modulation of the vertically- and the
horizontally-polarized components is roughly 20ns. If the modulating
phase changed within this 20ns, the QBER would be between a few
percent to 50\%, forming a waist. The waist in FIG. \ref{Fig:QBER}
has two points. This result is expected recalling that the step of
time shift is 10ns.


Our implementation is over 5km of telecom fibre. We address that
phase randomization itself does not limit the transmission distance.
It is the low intensity of the laser source (LD in FIG.
\ref{Fig:Schematic}) in our system that limits the transmission
distance \cite{PMLoss}. Transmission distance can be easily extended
by using a brighter laser diode.

In summary, we have performed the first QKD experiment with reliable
active phase randomization. Our result shows the global phase of
quantum signal is uniformly random. An important assumption in many
QKD security proofs
--- phase randomization --- is thus implemented with confidence. A potential security loophole is blocked. We
 expect phase randomization to become a standard part in future
QKD systems due to its significance in security \cite{ILM, GLLP,
LoPhase} and its feasibility.

We thank generous help from many colleagues including Li Qian and
Gr\'{e}goire Ribordy. Support of the funding agencies CFI, CIPI, the
CRC program, CIAR, MITACS, NSERC, OIT, and PREA is gratefully
acknowledged. This research was supported by Perimeter Institute for
Theoretical Physics. Research at Perimeter Institute is supported in
part by the Government of Canada through NSERC and by the province
of Ontario through MEDT.


\begin{thebibliography}{00}
\bibitem{PMLoss}
The extra phase modulator in our system introduces 12dB loss. This
high loss makes the synchronization system in Alice's side (mainly
PD in Fig. \ref{Fig:Schematic}) unstable. Another solution is that
we can use Alice's phase modulator ($\Phi_A$ in Fig.
\ref{Fig:Schematic}) to randomize the phase while encoding.
\end{thebibliography}
\end{document}